# Considerations of the duality in generalized coherent states formalism

**Dušan POPOV**
**University Politehnica Timisoara, Department of Physical Foundations of Engineering, Bd. V. Parvan No.2, 300223 Timisoara, Romania**
**dusan_popov@yahoo.co.uk**
**ORCID:** https://orcid.org/0000-0003-3631-3247.

*Abstract.* In the paper, a pair of dual operators is introduced with which a dual pair of generalized displacement operators is constructed. With these entities it is shown that the Barut - Girardello and Klauder - Perelomov generalized coherent states are dual states. The characteristics of these coherent states are constructed, separately and comparatively.

## 1. Introduction

The concept of coherent states (CSs) was introduced by Schrodinger, in 1926, for the linear harmonic oscillator (HO-1D) as the specific quantum state being often described as a state which has dynamics most closely resembling the oscillatory behavior of a classical harmonic oscillator [Schrodinger, 1926]. Although a few decades later the concept did not seem to have aroused much interest in specialized literature, in the middle of the last century physicists began to pay more attention to this notion and its specific applications, for a wide scientific field, starting with tha mathematical and solid state physics and up to the cosmology. At the same time, scientific interest and research were not limited to the coherent states and squeezed states of HO-1D, but they were also extended to other quantum systems so called nonlinear. This is how the concept of generalized or nonlinear coherent states (NCSs) appeared, based on the deformation of the bosonic ladder operators (the annihilation $a$ and creation $a^+$ canonical operators for HO-1D, as well as the number operator $\hat{n} = a^+ a$ ). As a consequence, CSs of HO-1D, also called canonical coherent states, were automatically considered to be linear.

For the case of HO-1D, the CSs can be derived from four definitions:

1.) The CSs are eigenstates of the annihilation operator (also called Barut-Girardello CSs) [Barut, 1971],

$$a\,|\,z>=z|\,z> \tag{1.1}$$

with $z = |\,z\,|\exp(\mathrm{i}\varphi)$ the complex variable labeling the CSs, $0 \le |\,z\,| \le R_c \le \infty$, $0 \le \varphi \le 2\pi$, where $R_c$ is the convergence radius of the CSs normalized function.

2.) By the quantum groups considerations, i.e. by applying the displacement operator $\hat{D}(z) = \exp(z\,a^+ - z^* a)$ on the HO-1D vacuum state (the Klauder-Perelomov manner) [Glauber, 1963], [Klauder, 1963], [Perelomov, 1972],

$$| z >= \hat{D}(z) | 0 > \tag{1.2}$$

3.) The CSs are the states with minimum value of uncertainty relation for the position $\hat{q}$ and momentum $\hat{p}$ operators

$$< \Delta\hat{q} >_z < \Delta\hat{p} >_z = \frac{1}{2} \quad , \quad \hbar = 1 \tag{1.3}$$

4.) The CSs form an overcomplete set satisfying the completeness relation, which often is called the unity operator expansion

$$\int \frac{d^2 z}{\pi} | z >< z | = \hat{I} = \sum_{n=0}^{\infty} | n >< n | \tag{1.4}$$

As is well known from quantum optics books (see. e.g. [Walls, 1995], [Vogel, 2006]), for the linear harmonic oscillator HO-1D the two ways of defining the coherent states (BG-, respectively KP-) lead to the same result:

$$| z >_{BG} = | z >_{KP} \equiv | z >_{HO-1D} = \exp\left( -\frac{1}{2} | z |^2 \right) \sum_{n=0}^{\infty} \frac{z^n}{\sqrt{n!}} | n > = \exp\left( -\frac{1}{2} | z |^2 + z a^+ \right) | 0 > \tag{1.5}$$

Also, and for this reason, these CSs are also called *linear*.

These states are normalized but not orthogonal, and constitute an overcomplete basis in the Hilbert space. At the same time, the CSs must accomplish the continuity condition in the label variable $z$.

However, in the case of other quantum systems (called for this reason, nonlinear), these definitions are not convergent, leading to different expressions for generalized CSs. The problem is to find the reason for this non-convergence. Alternatively, to find a pair of ladder operators that leads to the same type of CSs, regardless of their definition.

Between the two types of NCSs, that is Barut-Girardello (BG-CSs) and Klauder-Perelomov (KP-CSs), not long ago it was established that there is a certain duality [Roknizadeh, Tavassoly, 2004]. The aim of present paper is to examine the properties of the duality of two kind of NCSs, properties which are not appearing in the literature to our knowledge so far. Apart from the pure CSs we focused our attention on the mixed (thermal) states the subject which, to our knowledge, also has not appearing in the literature.

The generalized NCSs was defined by using the generalized annihilation $A_-$ and creation $A_+$ operators. The generalized or deformed annihilation operator is defined as a product between the canonical annihilation operator $a$ and a nonlinear function $f(\hat{N})$ depending on the operator number of particles $\hat{N} | n >= n | n >$, i.e. $A_- = a\, f(\hat{N})$. So, the NCSs are defined as the eigenvectors of the deformed annihilation operator. Similar, their conjugate, the generalized creation operator is $A_+ = (A_-)^+ = f^+(\hat{N}) a^+$ [Matos Filho, 1996], [Man'ko, 1997]. The function $f(\hat{N})$ is called the nonlinearity function and it can be complex or real. Because their phase is irrelevant, one may choose to be real and nonnegative, i.e. $f^+(\hat{N}) = f(\hat{N})$ .

In our analysis, we exhaustively used the properties of the generalized hypergeometric functions and the G-Meijer functions, which had the consequence of making the calculations easier, as well as checking their correctness.

The topic we are dealing with was previously examined, from a similar point of view, among others, by Roknizadeh and Tavassoly [Ali, 2004], [Roknizadeh, 2005], as well as by [Abbasi, 2010]. In this sense, the present paper takes a new and complementary look at the problem of duality in the formalism of generalized coherent states.

## 2. Duality of ladder operators $A_+$ and $\tilde{A}_-$

Let we consider a pair of annihilation and creation operators $A_-$ and $A_+$, whose action on the Fock vectors $|n>, \; n = 0,1,2,...$ are

$$A_- \,|\, n >= \sqrt{e(n)} \;|\, n-1> \quad , \quad A_+ \,|\, n >= \sqrt{e(n+1)} \;|\, n+1> \quad , \quad A_+ A_- \,|\, n >= e(n) \,|\, n> \tag{2.1}$$

If we choose $e(n) \equiv n\left[f(n)\right]^2$, where $f(n)$, with $f(0)=1$ , is the eigenvalue of the *nonlinearity function* $f(\hat{N})$, the above relations becomes

$$\begin{gathered} A_- \,|\, n >= \sqrt{n} \; f(n) \,|\, n-1> \quad , \quad A_+ \,|\, n >= \sqrt{n+1}\, f(n+1) |\, n+1> \quad , \\ A_+ A_- \,|\, n >= n\left[f(n)\right]^2 |\, n >= e(n) |\, n> \end{gathered} \tag{2.2}$$

For a real nonlinearity function $f(\hat{N}) = f^+(\hat{N})$, where $\hat{N} \,|\, n >= n \,|\, n>$ is the number operator, is valid the following commutation relation

$$\left[A_- \, , \, A_+ \,\right] = (\hat{N}+1)\left[f(\hat{N}+1)\right]^2 - \hat{N}\left[f(\hat{N})\right]^2 \neq 0 \tag{2.3}$$

which show that *generally,* the pair of annihilation and creation operators $A_-$ and $A_+$ *do not commute.* In a special case of linear CSs, in which case $f(\hat{N}) = 1$, the two operators are the standard bosonic canonically conjugate operators, i.e. $\left[A_- \, , \, A_+ \,\right] = 1$.

To avoid this situation, Roy and Roy defined a new pair of conjugate canonical operators, $B_-$ and $B_+$, which, together with the pair $A_-$ and $A_+$, satisfy the canonical commutation relations and allow the construction of two generalized displacement operators [Roy and Roy, 2000]:

Following Roy and Roy's idea, with some little modifications, let we consider two pairs of adjoints ladder operators, annihilation and creation, $A_\pm$ and $\tilde{A}_\pm$ , whose action on the Fock vectors $|n>, \; n = 0,1,2,...$ are

$$A_\pm \,|\, n >= \sqrt{e\left(n + \frac{1}{2} \pm \frac{1}{2}\right)} \;|\, n \pm 1> \qquad , \qquad \tilde{A}_\pm \,|\, n >= \sqrt{\tilde{e}\left(n + \frac{1}{2} \pm \frac{1}{2}\right)} \;|\, n \pm 1> \tag{2.4}$$

$$A_+ A_- \,|\, n >= e(n) \,|\, n> \qquad , \qquad \tilde{A}_+ \tilde{A}_- \,|\, n >= \tilde{e}(n) \,|\, n> \tag{2.5}$$

$$A_+ = \left(A_-\right)^+ , \quad A_- = \left(A_+\right)^+ , \quad \tilde{A}_+ = \left(\tilde{A}_-\right)^+ , \quad \tilde{A}_- = \left(\tilde{A}_+\right)^+ , \quad \tilde{\tilde{A}}_- = A_- \; , \quad \tilde{\tilde{A}}_+ = A_+ \tag{2.6}$$

We will use the "tilde" sign over the operators to denote their corresponding *duality property*. But, in what follows, for better clarity, for the different coherent states we will give up the "tilde" and use the corresponding lower index.

It is observed that the "tilde" operation differs from the hermetic conjugation operation, in the sense that it does not change the order of the operators in a product. Instead, it is similar to the simple operation of complex conjugation of numbers.

$$(A_+\widetilde{A}_-\widetilde{)} = \widetilde{A}_+ A_- \quad , \qquad (\widetilde{A}_- A_+\widetilde{)} = A_-\widetilde{A}_+ \tag{2.7}$$

In principle, instead of the basis of the Fock vectors $\{|\,n>, \ n = 0,1,2,...\}$, any orthonormal basis, e.g. $\{\Phi_n(x), \ n = 0,1,2,...\}$, can be used.

Let's choose the expression for $e(n)$ and $\widetilde{e}(n)$ in the following manner:

$$e(n) = n\,\frac{\prod_{j=1}^{q}\left(b_j - 1 + n\right)}{\prod_{i=1}^{p}\left(a_i - 1 + n\right)} \equiv n\,\frac{\left(\boldsymbol{b}_j - 1 + n\right)}{\left(\boldsymbol{a}_i - 1 + n\right)} \ , \qquad \widetilde{e}(n) = n\,\frac{\prod_{i=1}^{p}\left(a_i - 1 + n\right)}{\prod_{j=1}^{q}\left(b_j - 1 + n\right)} \equiv n\,\frac{\left(\boldsymbol{a}_i - 1 + n\right)}{\left(\boldsymbol{b}_j - 1 + n\right)} \tag{2.8}$$

where $p$ and $q$ are positive integers, and we used a short notation for the sequence of real numbers is $\boldsymbol{a} \equiv \{a_1, a_2, ..., a_p\}$ and $\boldsymbol{b} \equiv \{b_1, b_2, ..., b_q\}$.

This choice of special form is not random. Her motivation will become evident in what follows and will lead to normalization functions of CSs which will be generalized hypergeometric functions.

The eigenvalues of the products $A_+\widetilde{A}_-$ and $\widetilde{A}_- A_+$, respectively their “tilda conjugate” are

$$A_+\widetilde{A}_-\,|\,n> = A_+\sqrt{\widetilde{e}(n)}\,|\,n-1> = \sqrt{\widetilde{e}(n)}\sqrt{e(n)}\,|\,n> = \sqrt{n\,\frac{\left(\boldsymbol{a}_i - 1 + n\right)}{\left(\boldsymbol{b}_j - 1 + n\right)}}\sqrt{n\,\frac{\left(\boldsymbol{b}_j - 1 + n\right)}{\left(\boldsymbol{a}_i - 1 + n\right)}}\,|\,n> = n|\,n> \tag{2.9}$$

$$\left\{\begin{matrix}\widetilde{A}_- A_+ \\ A_-\widetilde{A}_+\end{matrix}\right\}|\,n> = (n+1)|\,n> \quad , \qquad \left\{\begin{matrix}\widetilde{A}_+ A_- \\ A_+\widetilde{A}_-\end{matrix}\right\}|\,n> = n|\,n> \tag{2.10}$$

Appealing to the well-known expression for the particle number operator, $\hat{\boldsymbol{N}}\,|\,n> = n\,|\,n>$, we can make the identification, $\widetilde{A}_+ A_- = A_+\widetilde{A}_- = \hat{\boldsymbol{N}}$ and $\widetilde{A}_- A_+ = A_-\widetilde{A}_+ = \hat{\boldsymbol{N}} + 1$, so that the following commutation expressions are valid

$$\left[\widetilde{A}_-\,,\,A_+\right] = 1 \quad , \qquad \left[A_-\,,\,\widetilde{A}_+\right] = 1 \tag{2.11}$$

$$\left[N\,,\,A_+\right] = +A_+ \quad , \qquad \left[N\,,\,\widetilde{A}_+\right] = +\widetilde{A}_+ \tag{2.12}$$

$$\left[N\,,\,\widetilde{A}_-\right] = -\widetilde{A}_- \quad , \qquad \left[N\,,\,A_-\right] = -A_- \quad . \tag{2.13}$$

These relations show that the operators dual sets of operators $\tilde{A}_-$ and $A_+$, respectively $A_-$ and $\tilde{A}_+$ are canonically conjugated. On the other hand, the set of operators $\left(\tilde{A}_-\ ,\ A_+\ ,\ \hat{N}\right)$ respectively their counterparts $\left(A_-\ ,\ \tilde{A}_+,\ \hat{N}\right)$, follow a Lie algebra. The algebra defined by the above equations is sometimes called the Heisenberg-Weyl algebra.

Let's define two new discrete strictly positive parameter functions $\rho(n)$, and $\tilde{\rho}(n)$ depending on the main quantum number $n$, also called the structure functions (as will be seen in the chapter on coherent states):

$$\rho(n) \equiv \prod_{s=1}^{n} e(s) = n!\frac{\prod_{j=1}^{q}\left(b_j\right)_n}{\prod_{i=1}^{p}\left(a_i\right)_n} \equiv n!\frac{\prod \boldsymbol{a}_i}{\prod \boldsymbol{b}_j} \quad , \qquad \tilde{\rho}(n) \equiv \prod_{s=1}^{n} \tilde{e}(s) = n!\frac{\prod_{i=1}^{p}\left(a_i\right)_n}{\prod_{j=1}^{q}\left(b_j\right)_n} \equiv n!\frac{\prod \boldsymbol{b}_j}{\prod \boldsymbol{a}_i} \tag{2.14}$$

such as $\rho(n)\,\tilde{\rho}(n) = (n!)^2$.

Therefore, here and in the following, the entities $x$ and $\tilde{x}$ can be considered as being dual.

Using the definition of the dual operators $A_\pm$ and $\tilde{A}_\pm$, we obtain

$$\begin{gathered}
|\,n> = \frac{1}{\sqrt{\rho(n)}}\left(A_+\right)^n |\,0>\ ,\ \ <n\,| = \frac{1}{\sqrt{\rho(n)}} <0\,|\left(A_-\right)^n\ , \\
|\,n> = \frac{1}{\sqrt{\tilde{\rho}(n)}}\left(\tilde{A}_+\right)^n |\,0>\ ,\ \ <n\,| = \frac{1}{\sqrt{\tilde{\rho}(n)}} <0\,|\left(\tilde{A}_-\right)^n\ , \\
|\,n><n\,| = \frac{1}{\rho(n)}\left(A_+\right)^n |\,0><0\,|\left(A_-\right)^n = \frac{1}{\tilde{\rho}(n)}\left(\tilde{A}_+\right)^n |\,0><0\,|\left(\tilde{A}_-\right)^n
\end{gathered} \tag{2.15}$$

These relations will be useful in the coherent states formalism that follows.

## 3. DOOT as the generalization of IWOP

It is well known that the ladder quantum canonical bosonic operators, annihilation $a$ and creation $a^+$, used in quantum optics can be arranged in three forms of ordered products: 1. Normal ordering $:(a^+)^n a^m:_N \equiv (a^+)^n a^m$, i.e. $a^+$ is to left and $a$ is to right. Consequently, we have $<z\,|:G(a^+,a):_N|\,z> = G(z^*,z)$ ; 2. Anti-Normal ordering $:(a^+)^n a^m:_{AN} \equiv a^m (a^+)^n$, where we have $<z\,|:G(a,a^+):_{AN}|\,z> = G(z,z^*)$; and 3. Weyl ordering (or symmetric ordering) $:(a^+)^n a^m:_W \equiv \begin{pmatrix} n+m \\ n \end{pmatrix}^{-1} \times (\text{sum of all symmetric products of } n\,a^+ \text{ and } m\,a)$.

These three situations can be written as a formula of "s-order power-series expansions" [Fujii, 2004]

$$\left\{(a^{+})^{n} a^{m}\right\}_{s} = \sum_{k=0}^{\min\{n,m\}} k! \binom{n}{k} \binom{m}{k} \left(\frac{t-s}{2}\right)^{k} \left\{(a^{+})^{n-k} a^{m-k}\right\}_{t} \tag{3.1}$$

where

$$\left\{(a^{+})^{n} a^{m}\right\}_{s} \equiv \frac{\partial^{n+m}}{\partial z^{n}\, \partial(-z^{*})^{m}} \exp\left(z a^{+} - z^{*} a\right) \exp\left(\frac{1}{2} s |z|^{2}\right)\Bigg|_{z=0} \tag{3.2}$$

and the indices $s$ and $t$ are $+1$ (for normal), $0$ (for Weyl) and $-1$ (for anti-normal) operator ordering.

In order to unify the approach for arranging quantum operators of optical fields into ordered products, Hong-yi Fan proposed an *integration technique within an ordered product of operators*, which he named, for short, IWOP (see, the review article [Hong-yi Fan, 2003] and references therein). This technique refers *only* to the canonical bosonic operators annihilation $a$ and creation $a^{+}$, associated with the linear quantum harmonic oscillator (HO-1D).

A few years ago we introduced a new approach of normal ordering operator's products connected with the generalized hypergeometric CSs, called the *diagonal ordering operation technique* (DOOT) and denoted it with the symbol # # [Popov, 2015]. This ordering technique is applied *to any pair* of raising and lowering nonlinear operators $A_{+}$ and $A_{-}$, respectively $\tilde{A}_{+}$ and $\tilde{A}_{-}$, being a generalization of the integration within an ordered product (IWOP), introduced by Hong-yi Fan [Fan, 1999]. Given that the IWOP is applicable *only for Bose operators*, referring only to the CSs of the HO-1D, and DOOT is applicable to *any pair* of raising and lowering nonlinear operators, in this way the new formalism, called *diagonal ordering operation technique* (DOOT), contains IWOP as a special case.

We generalized the IWOP formalism of Fan by extending it for any pair of dual operators $A_{-}$ and $\tilde{A}_{+}$ , *as well as* $\tilde{A}_{-}$ and $A_{+}$.

The most important rules of the DOOT are (the rule for the pair $\tilde{A}_{-}$ and $\tilde{A}_{+}$ are the same):

a) The order of operators $A_{-}$ and $A_{+}$ can be permuted inside the symbol # #, so that finally we obtain a function of normally ordered operator product $\# f\left(A_{+} A_{-}\right) \#$, $\#(A_{-})^{n}(A_{+})^{n}\# = \#(A_{+})^{n}(A_{-})^{n}\# = (A_{+}A_{-})^{n}$ ;

b) inside the symbol # # we can perform all algebraic operations, according to the usual rules;

c) the operators $A_{-}$ and $A_{+}$ can be treated as simple *c*-numbers;

d) the vacuum state projector $|0\rangle\langle 0|$, in the frame of DOOT, has the following normal ordered form:

$$
\begin{aligned}
|0><0| &= \frac{1}{\sum_{n=0}^{\infty}\frac{1}{\rho(n)}\#\left(A_+A_-\right)^n\#} = \frac{1}{\#_pF_q(\boldsymbol{a};\boldsymbol{b};A_+A_-)\#} = \\
&= \frac{1}{\sum_{n=0}^{\infty}\frac{1}{\tilde{\rho}(n)}\#\left(\tilde{A}_+\tilde{A}_-\right)^n\#} = \frac{1}{\#_qF_p(\boldsymbol{b};\boldsymbol{a};\tilde{A}_+\tilde{A})\#}
\end{aligned}
\tag{3.3}
$$

where in the denominator appear an normal ordered operator function depending on the ordered operator product $\#A_+A_-\#$ as “argument”. In fact, it is a generalized hypergeometric function, as we will see below, defined as

$$
{}_pF_q(\boldsymbol{a};\boldsymbol{b};x) = \sum_{n=0}^{\infty}\frac{\prod_{i=1}^{p}(a_i)_n}{\prod_{j=1}^{q}(b_j)_n}\frac{x^n}{n!} \equiv \sum_{n=0}^{\infty}\frac{1}{\rho(n)}x^n
\tag{3.4}
$$

The motivation of the above equality is simple: starting from the completeness relation for the Fock vectors $\sum_{n=0}^{\infty}|n><n|=1$ and using the results of the multiple actions of operators $A_-$ and $A_+$ on the fundamental (vacuum) state vectors, the above relation is reached.

The function ${}_pF_q(\boldsymbol{a};\boldsymbol{b};A_+A_-)$ converges in the following cases [Won Sang Chung, 2014]:

| ***For*** | ***If*** | ***Condition*** |
|---|---|---|
| any $z$ | $p<q+1$ | - |
| $\|z\|<1$ | $p=q+1$ | - |
| $\|z\|=1$ | $p=q+1$ | $\eta=1$ |
| $\|z\|=1$ | $p=q+1$ | $0\le\eta<1$ |
| $z\neq 1$ | $p=q+1$ | $0\le\eta<1$ |

where $\eta=\mathrm{Re}\left(\sum_{i=1}^{p}a_i-\sum_{j=1}^{q}b_j\right)$.

As we mentioned, *the same rules of the DOOT are also valid for the pair of operators $\tilde{A}_-$ and $\tilde{A}_+$.*

Calculating the vacuum operator for pairs of dual operators $\left(A_+,\ \tilde{A}_-\right)$, respectively $\left(\tilde{A}_+,\ A_-\right)$, expressions different from those above are obtained:

$$
|n><n| = \frac{1}{\sqrt{\rho(n)}}\#\left(A_+\right)^n|0><0|\left(\tilde{A}_-\right)^n\#\frac{1}{\sqrt{\tilde{\rho}(n)}} = \frac{1}{n!}\#\left(A_+\right)^n|0><0|\left(\tilde{A}_-\right)^n\#
\tag{3.5}
$$

$$
\sum_n|n><n| = |0><0|\sum_n\frac{1}{n!}\#\left(A_+\tilde{A}_-\right)^n\# = 1
\tag{3.6}
$$

$$|0><0|=\#\exp\left(-A_{+}\tilde{A}_{-}\right)\# \tag{3.7}$$

and similar,

$$|0><0|=\#\exp\left(-\tilde{A}_{+}A_{-}\right)\# \tag{3.8}$$

The DOOT approach proved useful in studying the properties of CSs related to different quantum systems of oscillators: pseudoharmonics, Morse, Meixner, but also for deducing integrals in which the hypergeometric and Meijer functions are involved [Popov, 2022].

## 4. Generalized CSs

The choice of the above expression for the $e(n)$, and, as a consequence, for $\rho(n)$, which play the role of nonlinearity function is not accidental: it is closely related to the definition of generalized hypergeometric function ${}_pF_q(\boldsymbol{a};\boldsymbol{b};x)$.

We used the Pochhammer symbols, with following properties, for $n=1,2,3,\ ...$:

$$(a)_n=\frac{\Gamma(a+n)}{\Gamma(a)}=a(a+1)...\,(a+k-1),\ (0)_0=1,\ (0)_n=0,\ (a)_0=1 \tag{4.1}$$

The generalized hypergeometric function can be represented through Meijer G-function

$${}_pF_q(\boldsymbol{a};\boldsymbol{b};x)=\frac{\prod_{j=1}^{q}\Gamma(b_j)}{\prod_{i=1}^{p}\Gamma(a_i)}G_{p,q+1}^{1,p}\left(-x\ \middle|\ \begin{matrix}\boldsymbol{1-a}\ ; & / \\ 0\ ; & \boldsymbol{1-b}\end{matrix}\right);\quad \Gamma(\boldsymbol{b}\,/\,\boldsymbol{a})\equiv\frac{\prod_{j=1}^{q}\Gamma(b_j)}{\prod_{i=1}^{p}\Gamma(a_i)} \tag{4.2}$$

The Meijer's G-function satisfies the following classical integral [Mathai, 1973]:

$$\int_0^\infty dx\, x^{s-1}\, G_{p,q}^{m,n}\left(\beta x\ \middle|\ \begin{matrix}\{a_i\}_1^n\ ; & \{a_i\}_{n+1}^p \\ \{b_j\}_1^m\ ; & \{b_j\}_{m+1}^q\end{matrix}\right)=\frac{1}{\beta^s}\frac{\prod_{j=1}^{m}\Gamma(b_j+s)\prod_{i=1}^{n}\Gamma(1-a_i-s)}{\prod_{j=m+1}^{q}\Gamma(1-b_j-s)\prod_{i=n+1}^{p}\Gamma(a_i+s)} \tag{4.3}$$

Generally, there are following definitions or approaches to generalized or nonlinear coherent states (NCSs):

- The Barut-Girardello coherent states (BG-CSs), defined as eigenstates of the nonlinear annihilation operator [Barut, 1971].
- The Klauder-Perelomov coherent states (KP-CSs), defined through group-theoretical approach [Perelomov, 1972].
- The Gazeau-Klauder coherent states (GK-CSs), which are nonspreading and temporally stable, directly related to the Hamiltonian of the examined system [Gazeau, 1999]. The GK-CSs can be obtained by acting with the exponential operator $\#\exp\left(-\mathrm{i}\gamma A_+ A_-\right)\#$ on the non-normalized BG-CSs depending on the real positive variable $|J>$ and considering the rules of DOOT (for details see e.g. [Popov, 2016]).

In the present paper we will deal only with the first two types of NCSs, which we will consider as a *dual pair* and highlight the similarities and differences between their properties.

The NCSs, regardless of its type (definition), have the following expansion according to the expansion in the Fock vectors-basis:

$$|z> = \frac{1}{\sqrt{\mathcal{N}\left(|z|^2\right)}} \sum_{n=0}^{\infty} \frac{z^n}{\sqrt{\rho(n)}} |n> = \frac{1}{\sqrt{\mathcal{N}\left(|z|^2\right)}} \mathcal{N}\left(zA_+\right)|0> \quad , \tag{4.4}$$

and where the normalization function $\mathcal{N}\left(|z|^2\right)$ is obtained from the normalization condition $<z|z>=1$ :

$$\mathcal{N}\left(|z|^2\right) = \sum_{n=0}^{\infty} \frac{1}{\rho(n)} \left(|z|^2\right)^n \quad . \tag{4.5}$$

For the states to belong to the Fock space, the condition must be fulfilled $0 < \mathcal{N}\left(|z|^2\right) < \infty$ or, in other words, the convergence radius $R_c$ of the series to fulfill the following inequality: $0 < |z| < R_c = \lim_{n\to\infty} e(n)$ .

The *minimal conditions* that the set of NCSs must fulfill were formulated by Klauder and that is why they are also called "Klauder's minimal prescriptions" [Klauder, 1963]:

(I). NCSs are normalized but non-orthogonal:

$$<z|z'> = \frac{\mathcal{N}\left(z^* z'\right)}{\sqrt{\mathcal{N}\left(|z|^2\right)}\sqrt{\mathcal{N}\left(|z'|^2\right)}} = \begin{cases} 1, & z = z' \\ \neq 0, & z \neq z' \end{cases} \tag{4.6}$$

but they form an overcomplete set.

(II). They must be continuous in the label variable $z$ :

$$\lim_{z\to z'} \| z - z' \| = \lim_{\substack{r\to r' \\ \varphi\to\varphi'}} \sqrt{r^2 + r'^2 - 2rr'\cos(\varphi - \varphi')} = 0 \tag{4.7}$$

(III). It must be necessary to satisfy the resolution of the identity, i.e. to close a resolution of the identity

$$\int d\mu(z)|z><z| = \hat{I} = \sum_{n=0}^{\infty} |n><n| \tag{4.8}$$

with the integration measure $d\mu(z) = \frac{d^2 z}{\pi} = \frac{d\varphi}{2\pi} d\left(|z|^2\right) h\left(|z|^2\right)$ and weight function $h\left(|z|^2\right)$ which must be found for each individual case.

Inserting in the resolution of the identity the expressions of integration measure as well as for the expansion of NCSs, in order to be satisfied the completeness relation $\sum_{n=0}^{\infty} | n >< n | = 1$ for the Fock vectors, after the angular integration

$$\int_0^{2\pi} \frac{d\varphi}{2\pi} \left( z^* \right)^n z^{n'} = \left( | z |^2 \right)^n \delta_{n n'} \tag{4.9}$$

as well as the function change $\tilde{h}\left( | z |^2 \right) = \frac{h\left( | z |^2 \right)}{\mathcal{N}\left( | z |^2 \right)}$ and exponent change $n = s - 1$ we have to solve the following moment problem in variable $| z |^2$:

$$\int_0^{R_c} d\left( | z |^2 \right) \tilde{h}\left( | z |^2 \right) \left( | z |^2 \right)^{s-1} = \rho(s-1) \tag{4.10}$$

Depending on the value of convergence radius $R_c$ (which is calculable using one of the convergence criteria of the power series, for example the ratio criterion), we can have two situations [Kla-Pen-Six]:

$$R_c = \lim_{n \to \infty} \frac{\rho(n)}{\rho(n+1)} = \begin{cases} \infty, & \text{Stieltjes moment problem} (= \text{SM}) \\ < \infty, & \text{Hausdorff moment problem} (= \text{HM}) \end{cases} \tag{4.11}$$

In a compact and unitary expression, the moment problem is written as [Roknizadeh, 2004]

$$\int_0^{\infty} d\left( | z |^2 \right) \begin{pmatrix} \tilde{h}\left( | z |^2 \right) \\ H\left( R_c - | z |^2 \right) \tilde{h}\left( | z |^2 \right) \end{pmatrix} \left( | z |^2 \right)^{s-1} = \rho(s-1) \qquad \leftrightarrow \qquad \begin{pmatrix} SM \\ HM \end{pmatrix} \tag{4.12}$$

where $H\left( R_c - | z |^2 \right)$ is the Heaviside step function:

$$H\left( R_c - | z |^2 \right) = \begin{cases} 0, & | z |^2 > R_c \\ 1, & | z |^2 \leq R_c \end{cases} \tag{4.13}$$

Generally, the solution $\tilde{h}\left( | z |^2 \right)$ is proportional to a Meijer G-function and its concrete form depends on the type of coherent state. Consequently, the weight function, which must necessarily be positive, is

$$h\left( | z |^2 \right) = C(\boldsymbol{a}, \boldsymbol{b}) \ \mathcal{N}\left( | z |^2 \right) G_{p,q}^{m,n}\left( | z |^2 \left| \begin{matrix} \{a_i\}_1^n \ ; & \{a_i\}_{n+1}^p \\ \{b_j\}_1^m \ ; & \{b_j\}_{m+1}^q \end{matrix} \right. \right) \tag{4.14}$$

where $C(\boldsymbol{a}, \boldsymbol{b})$ is a constant that depends on the internal structure of the nonlinearity function $f(n)$.

Finally, the integration measure becomes:

$$d\mu(z) = C(\boldsymbol{a}, \boldsymbol{b}) \frac{d\varphi}{2\pi} d\left( | z |^2 \right) \mathcal{N}\left( | z |^2 \right) G_{p,q}^{m,n}\left( | z |^2 \left| \begin{matrix} \{a_i\}_1^n \ ; & \{a_i\}_{n+1}^p \\ \{b_j\}_1^m \ ; & \{b_j\}_{m+1}^q \end{matrix} \right. \right) \tag{4.15}$$

Generally, in the frame of DOOT, the projector of NCSs can be written as

$$|\, z><z\,| = \frac{1}{\mathcal{N}(|\, z\,|^2)} \#\mathcal{N}(zA_+)|\,0><0\,|\mathcal{N}(z^* A_-)\# = \frac{1}{\mathcal{N}(|\, z\,|^2)} \# \frac{\mathcal{N}(zA_+)\mathcal{N}(z^* A_-)}{\mathcal{N}(A_+ A_-)} \# \qquad (4.16)$$

As a result, the resolution of identity leads us to the condition

$$\int \frac{d^2 z}{\pi} G_{p,q}^{m,n}\left( |\, z\,|^2 \left| \begin{matrix} \{a_i\}_1^n \; ; & \{a_i\}_{n+1}^p \\ \{b_j\}_1^m \; ; & \{b_j\}_{m+1}^q \end{matrix} \right. \right) \#\mathcal{N}(zA_+)\mathcal{N}(z^* A_-)\# = \frac{1}{C(\boldsymbol{a},\boldsymbol{b})} \#\mathcal{N}(A_+ A_-)\# \qquad (4.17)$$

Essentially, this is a new generalized integral in which the operators are regarded as simple c-numbers, and consequently can be replaced by some constants.

After performing the angular integration

$$\int_0^{2\pi} \frac{d\varphi}{2\pi} \#\mathcal{N}(zA_+)\mathcal{N}(z^* A_-)\# = \sum_n \frac{\#(A_+ A_-)^n \#}{[\rho(n)]^2} \left(|\, z\,|^2\right)^n \qquad (4.18)$$

the above integral in real space becomes

$$\sum_n \frac{\#(A_+ A_-)^n \#}{[\rho(n)]^2} \int_0^\infty d\left(|\, z\,|^2\right)\left(|\, z\,|^2\right)^n G_{p,q}^{m,n}\left( |\, z\,|^2 \left| \begin{matrix} \{a_i\}_1^n \; ; & \{a_i\}_{n+1}^p \\ \{b_j\}_1^m \; ; & \{b_j\}_{m+1}^q \end{matrix} \right. \right) = \frac{1}{C(\boldsymbol{a},\boldsymbol{b})} \#\mathcal{N}(A_+ A_-)\# \qquad (4.19)$$

In order to obtain equality, the real integral must be

$$\int_0^\infty d\left(|\, z\,|^2\right)\left(|\, z\,|^2\right)^n G_{p,q}^{m,n}\left( |\, z\,|^2 \left| \begin{matrix} \{a_i\}_1^n \; ; & \{a_i\}_{n+1}^p \\ \{b_j\}_1^m \; ; & \{b_j\}_{m+1}^q \end{matrix} \right. \right) = \frac{1}{C(\boldsymbol{a},\boldsymbol{b})} \rho(n) \qquad (4.20)$$

*Statistical properties* of the NCSs are examined with the help of the expectation value $<\hat{\boldsymbol{N}}^s>_z \equiv <z\,|\,\hat{\boldsymbol{N}}^s\,|\,z>$ of an integer power of number operator $\hat{\boldsymbol{N}}$.

$$<\hat{\boldsymbol{N}}^s>_z = \frac{1}{\mathcal{N}(|\, z\,|^2)} \left(|\, z\,|^2 \; \hat{D}_{|z|^2}\right)^s \mathcal{N}(|\, z\,|^2) \qquad (4.21)$$

We need to compute the Mandel parameter, defined as [Walls, 1995].

$$Q_{|z|} = \frac{<\hat{\boldsymbol{N}}^2>_z - \left(<\hat{\boldsymbol{N}}>_z\right)^2}{<\hat{\boldsymbol{N}}>_z} - 1 = |\, z\,|^2 \left[ \frac{\left(\hat{D}_{|z|^2}\right)^2 \mathcal{N}(|\, z\,|^2)}{\hat{D}_{|z|^2} \mathcal{N}(|\, z\,|^2)} - \frac{\hat{D}_{|z|^2} \mathcal{N}(|\, z\,|^2)}{\mathcal{N}(|\, z\,|^2)} \right] \qquad (4.22)$$

Regarding the behavior of NCSs, i.e. the expression of their distribution function $\mathrm{P}_{|z|^2}^{(\ldots)}$, the following three situations can exist:

$$Q_{|z|} = \begin{cases} > 0, & \text{ordinary (classical) states, called super-Poissonian (bunching states)} \\ = 0, & \text{canonical states, Poissonian statistics or photon number distribution} \\ < 0, & \text{non-classical states, said to be sub-Poissonian (anti-bunching)} \end{cases} \qquad (4.23)$$

This behavior can be verified by comparing the distribution function $\mathrm{P}_{|z|^2}^{(\ldots)} \equiv |<z|n>|^2$ versus Poisson distribution $\mathrm{P}_{|z|^2}^{(Poisson)} = e^{-|z|^2} \frac{\left(|z|^2\right)^n}{n!}$, separately for BG- and KP-CSs. Then, we have the following situations:

| | |
|---|---|
| $\mathrm{P}_{\|z\|^2}^{(\ldots)} > \mathrm{P}_{\|z\|^2}^{(Poisson)}$ | super-Poissonian distribution |
| $\mathrm{P}_{\|z\|^2}^{(\ldots)} = \mathrm{P}_{\|z\|^2}^{(Poisson)}$ | Poissonian distribution |
| $\mathrm{P}_{\|z\|^2}^{(\ldots)} < \mathrm{P}_{\|z\|^2}^{(Poisson)}$ | sub-Poissonian distribution |

If we refer to the *mixed states*, the approach is quite different. As a characteristic example of mixed states we will refer to the thermal states of a quantum canonical ensemble at equilibrium temperature $T = 1/(k_B \beta_T)$, with energy eigenvalues $E_n$, whose equilibrium density operator $\hat{\rho}$ , and also the partition function $Z(\beta)$ are

$$\hat{\rho} = \frac{1}{Z(\beta_T)} \sum_n e^{-\beta_T E_n} |n><n| \qquad , \qquad Z(\beta_T) = \sum_n e^{-\beta_T E_n} \tag{4.24}$$

In the frame of NCSs, the ensemble is characterized, on the one hand, by the diagonal element of the density operator, i.e. by the *Q- distribution function (Husimi's function)*

$$Q(|z|^2) \equiv <z|\hat{\rho}|z> = \frac{1}{\mathcal{N}\left(|z|^2\right)} \sum_n e^{-\beta_T E_n} \frac{(|z|^2)^n}{\rho(n)} \tag{4.25}$$

and, on the other hand, by the *P- quasi distribution function*, which appear as the weight function in the diagonal expansion of density operator $\hat{\rho}$ with respect to the NCSs projector $|z><z|$:

$$\hat{\rho} = \int d\mu(z)\, P(|z|^2) |z><z| \tag{4.26}$$

Furthermore, the concrete expressions for $\hat{\rho}$, respectively the *Q*- and *P*- functions are different, due to the different expression of $e(n)$. For example, for a *linear energy spectrum*

$$E_n = \hbar\omega e(n) = \hbar\omega\left(n + e_0\right) \tag{4.27}$$

the density operator and partition function are

$$\hat{\rho} = \frac{e^{-\beta_T \hbar\omega\, e_0}}{Z(\beta_T)} \sum_{n=0}^{\infty} \left(e^{-\beta_T \hbar\omega}\right)^n |n><n| = \frac{1}{\bar{n}+1} \sum_{n=0}^{\infty} \left(\frac{\bar{n}}{\bar{n}+1}\right)^n |n><n| \quad , \tag{4.28}$$

$$Z(\beta_T) = e^{-\beta_T \hbar\omega\, e_0} \sum_{n=0}^{\infty} \left(e^{-\beta_T \hbar\omega}\right)^n = e^{-\beta_T \hbar\omega\, e_0} \frac{1}{e^{\beta_T \hbar\omega} - 1} = \left(\frac{\bar{n}}{\bar{n}+1}\right)^{e_0} \left(\bar{n}+1\right) \tag{4.29}$$

where $\bar{n} = \left(e^{\beta_T \hbar\omega} - 1\right)^{-1}$ is the *Bose-Einstein distribution function*.

For other more complicated energy spectra, specific methods must be applied. For example, for a quadratic spectrum (characteristic, e.g. for the Morse oscillator) a specific *ansatz* can be applied [Popov, 2003].

Generally, using the DOOT rules, the density operator becomes

$$\hat{\rho} = \frac{1}{Z(\beta_T)} \# \frac{1}{\mathcal{N}(A_+ A_-)} \sum_n e^{-\beta E_n} \frac{(A_+ A_-)^n}{\rho(n)} \# \tag{4.30}$$

with their diagonal expansion

$$\hat{\rho} = \frac{1}{Z(\beta_T)} C(\boldsymbol{a},\boldsymbol{b}) \# \frac{1}{\mathcal{N}(A_+ A_-)} \int_0^\infty d(|z|^2) \frac{h(|z|^2)}{\mathcal{N}(|z|^2)} P(|z|^2) \int_0^{2\pi} \frac{d\varphi}{2\pi} \mathcal{N}(zA_+) \mathcal{N}(z^* A_-) \# \tag{4.31}$$

The angular integral is

$$\# \int_0^{2\pi} \frac{d\varphi}{2\pi} \mathcal{N}(zA_+) \mathcal{N}(z^* A_-) \# = \sum_n \frac{\#(A_+ A_-)^n \#}{[\rho(n)]^2} (|z|^2)^n \tag{4.32}$$

so that we have

$$\hat{\rho} = \frac{1}{Z(\beta_T)} C(\boldsymbol{a},\boldsymbol{b}) \# \frac{1}{\mathcal{N}(A_+ A_-)} \sum_n \frac{(A_+ A_-)^n}{[\rho(n)]^2} \# \int_0^\infty d(|z|^2) \frac{h(|z|^2)}{\mathcal{N}(|z|^2)} P(|z|^2) (|z|^2)^n \tag{4.33}$$

If we perform the function change

$$h_{red}(|z|^2) \equiv \frac{h(|z|^2)}{\mathcal{N}(|z|^2)} P(|z|^2) \tag{4.34}$$

it is obvious that we will have to solve the following problem of moments, similar to the one before

$$\int_0^\infty d(|z|^2) h_{red}(|z|^2) (|z|^2)^n = \frac{1}{C(\boldsymbol{a},\boldsymbol{b})} e^{-\beta E_n} \rho(n) \tag{4.35}$$

Next, the calculation depends on the concrete expression of the energy eigenvectors $E_n$ for each quantum system examined.

The statistical behavior of the thermal states can be revealed if we calculate *the thermal counterpart of the Mandel parameter* $Q_{th}(\beta_T)$ (used in [Ghostal, 1995], [Popov, 2003], [Laforgia, 2010] ). Their meaning of its values is the same as for the Mandel parameter for pure NCSs. The thermal expectations being *independent on the representation*, it is possible to examine the thermal Mandel parameter together for both types of CSs, BG and KP:

$$Q_{\mathrm{th}}(\beta_T) = \frac{< \hat{\boldsymbol{N}}^2 >_{th} - \left(< \hat{\boldsymbol{N}} >\right)_{th}^2}{< \hat{\boldsymbol{N}} >_{th}} - 1 \tag{4.36}$$

where, the thermal averages of the integer powers of number operator are

$$< \hat{\boldsymbol{N}}^s >_{th} = \mathrm{Tr}\left(\hat{\rho} \hat{\boldsymbol{N}}^s\right) = \frac{1}{Z(\beta_T)} \sum_{n=0}^\infty n^s e^{-\beta_T E_n} \tag{4.37}$$

This leads to

$$Q_{\rm th}(\beta_T)=\frac{\sum_{n=0}^{\infty}n^2\,e^{-\beta_T E_n}}{\sum_{n=0}^{\infty}n\,e^{-\beta_T E_n}}-\frac{\sum_{n=0}^{\infty}n\,e^{-\beta_T E_n}}{\sum_{n=0}^{\infty}e^{-\beta_T E_n}}-1 \tag{4.38}$$

from where the value of the thermal Mandel parameter can be calculated for each quantum system separately. For example, for a system that has a linear energy spectrum $E_n=\hbar\omega e(n)=\hbar\omega(n+e_0)$, we have $<\hat{\boldsymbol{N}}>_{th}=\bar{n}$ and $<\hat{\boldsymbol{N}}^2>_{th}=\bar{n}+2(\bar{n})^2$ and it is obtained that the thermal Mandel parameter is even equal to the Bose-Einstein distribution function:

$$Q_{\rm th}(\beta_T)=\bar{n}=\frac{1}{e^{-\beta_T\hbar\omega}-1}>0 \tag{4.39}$$

This means that the mixed thermal states for the systems with a linear energy spectrum have a supra-Poissonian behavior for any temperature.

## 5. Barut-Girardello versus Klauder-Perelomov coherent states

Following a traditional way (one can even say "classical"), let's build the coherent states with the pair of dual operators $(A_-\ ,\ A_+)$ and $(\tilde{A}_-\ ,\ \tilde{A}_+)$, in the spirit of the founders of this concept (see, e.g. [Glauber, 1963], [Sudarshan, 1963], [Barut, 1971], [Perelomov, 1972], [Gilmore, 1974], [Klauder, 1985], and not only). First, we will construct CSs using the pair of operators $(A_-\ ,\ A_+)$**.** Next, we will do the same thing using the set of dual operators $(\tilde{A}_-\ ,\ \tilde{A}_+)$, in order to establish the connection between the *dual coherent states* BG-CSs and KP-CSs.

Following an idea formulated by B Roy and P Roy [Roy, 2000], with the help of the two pairs of dual operators we can construct two *dual generalized displacement operators*:

$$\mathcal{D}(z)=\exp\left(z\,A_+-z^*\tilde{A}_-\right)=\frac{1}{\sqrt{\mathcal{N}_1(|z|^2)}}\exp\left(z\,A_+\right)\exp\left(-z^*\tilde{A}_-\right) \tag{5.1}$$

$$\tilde{\mathcal{D}}(z)=\exp\left(z\,\tilde{A}_+-z^*A_-\right)=\frac{1}{\sqrt{\tilde{\mathcal{N}}_2(|z|^2)}}\exp\left(z\,\tilde{A}_+\right)\exp\left(-z^*A_-\right) \tag{5.2}$$

It is observed that these expressions can be obtained from each other through the "tilde conjugation" operation. The normalization functions $\mathcal{N}_1(|z|^2)$ and $\tilde{\mathcal{N}}_2(|z|^2)$ will be determined from the CSs normalization condition.

**Barut-Girradello nonlinear coherent states** (BG-CSs) are defined as eigenvectors of the nonlinear annihilation operator [Barut, 1971], [Roy, 2000]:

*Definition 1*, with the help of the annihilation operator $A_-$:

$$A_-\,|\,z>_{BG}=z|\,z>_{BG} \tag{5.3}$$

and their expansion according to the Fock vectors is

$$|\, z >_{BG} = \frac{1}{\sqrt{{}_pF_q(\boldsymbol{a};\, \boldsymbol{b}; |\, z\, |^2)}} \sum_{n=0}^{\infty} \frac{z^n}{\sqrt{\rho_{BG}(n)}} |\, n > = \frac{1}{\sqrt{{}_pF_q(\boldsymbol{a};\, \boldsymbol{b}; |\, z\, |^2)}}\, {}_pF_q(\boldsymbol{a};\, \boldsymbol{b}; |\, zA_+) |\, 0 > \tag{5.4}$$

where $\rho_{BG}(n) = \rho(n)$.

*Definition 2*, with the help of the generalized displacement operator $\tilde{\mathcal{D}}(z)$:

$$|\, z >_{BG} = \tilde{\mathcal{D}}(z) |\, 0 > \tag{5.5}$$

$$\begin{aligned} |\, z >_{BG} &= \frac{1}{\sqrt{\tilde{\mathcal{N}}_2(|z|^2)}} \exp\left(z \tilde{A}_+\right) \exp\left(-z^* A_-\right) |\, 0 > = \frac{1}{\sqrt{\tilde{\mathcal{N}}_2(|z|^2)}} \exp\left(z \tilde{A}_+\right) |\, 0 > = \\ &= \frac{1}{\sqrt{\tilde{\mathcal{N}}_2(|z|^2)}} \sum_n \frac{z^n}{n!} \left(\tilde{A}_+\right)^n |\, 0 > = \frac{1}{\sqrt{\tilde{\mathcal{N}}_2(|z|^2)}} \sum_n \frac{z^n}{n!} \sqrt{\tilde{\rho}(n)}\, |\, n > = \\ &= \frac{1}{\sqrt{{}_pF_q(\boldsymbol{a};\, \boldsymbol{b}; |\, z\, |^2)}} \sum_{n=0}^{\infty} \frac{z^n}{\sqrt{\rho_{BG}(n)}} |\, n > \end{aligned} \tag{5.6}$$

where $\tilde{\rho}(n) \equiv \rho_{KP}(n) = \dfrac{(n!)^2}{\rho_{BG}(n)}$ and, after normalization, $\tilde{\mathcal{N}}_2(|z|^2) = {}_pF_q(\boldsymbol{a};\, \boldsymbol{b}; |\, z\, |^2)$.

It is observed, therefore, that *both definitions lead to the same expression* of BG-CSs.

The convergence radius $R_{BG}$ can be calculated having in mind that $(x)_{n+1} = (x+n)(x)_n$:

$$R_{BG} = \lim_{n\to\infty} \frac{\rho_{BG}(n)}{\rho_{BG}(n+1)} = \lim_{n\to\infty} n^{p-q-1} \frac{1}{1+\frac{1}{n}} \frac{\prod_{i=1}^{p}\left(\frac{a_i}{n}+1\right)}{\prod_{j=1}^{q}\left(\frac{b_j}{n}+1\right)} = \lim_{n\to\infty} n^{p-q-1} = \begin{cases} \infty, & \text{if} \quad p-q-1>0 \\ 1\,, & \text{if} \quad p-q-1=0 \\ 0\,, & \text{if} \quad p-q-1<0 \end{cases} \tag{5.7}$$

and strongly depend on the values of the indices $p$ and $q$.

The BG-CSs projector is

$$|\, z >_{BG\; BG}< z\, | = \frac{1}{{}_pF_q(\boldsymbol{a};\, \boldsymbol{b}; |\, z\, |^2)} \# \frac{{}_pF_q(\boldsymbol{a};\, \boldsymbol{b}; |\, zA_+)\; {}_pF_q(\boldsymbol{a};\, \boldsymbol{b}; |\, z^* A_-)}{{}_pF_q(\boldsymbol{a};\, \boldsymbol{b}; A_+ A_-)} \# \tag{5.8}$$

and the non orthogonality relation becomes

$${}_{BG}< z\, |\, z' >_{BG} = \frac{{}_pF_q(\boldsymbol{a};\, \boldsymbol{b}; z^* z')}{\sqrt{{}_pF_q(\boldsymbol{a};\, \boldsymbol{b}; |\, z\, |^2)}\, \sqrt{{}_pF_q(\boldsymbol{a};\, \boldsymbol{b}; |\, z'|^2)}} \tag{5.9}$$

To ensure the decomposition of the unit operator, it is necessary to solve the following moment problem (where $z = s-1$):

$$\int_0^\infty d\left(|z|^2\right)\widetilde{h}\left(|z|^2\right)\left(|z|^2\right)^{s-1} == \rho_{BG}(n) = \Gamma\left(\boldsymbol{a}/\boldsymbol{b}\right)\Gamma(s)\,\frac{\Gamma(s)\prod_{j=1}^{q}\Gamma\left(b_j - 1 + s\right)}{\prod_{i=1}^{p}\Gamma\left(a_i - 1 + s\right)} \tag{5.10}$$

which leads to the following expression of the integration measure:

$$d\mu_{BG}(z) = \Gamma\left(\boldsymbol{a}/\boldsymbol{b}\right)\frac{d\varphi}{2\pi}d(|z|^2)\ {}_pF_q(\boldsymbol{a};\,\boldsymbol{b};|z|^2)\ G_{p,\,q+1}^{q+1,0}\left(|z|^2\left|\begin{matrix}/\,; & \boldsymbol{a-1}\\ 0\,,\ \boldsymbol{b-1}\,; & /\end{matrix}\right.\right) \tag{5.11}$$

Consequently, the resolution of identity becomes

$$\int\frac{d^2z}{\pi}G_{p,q+1}^{q+1,0}\left(|z|^2\left|\begin{matrix}/\,; & \boldsymbol{a-1}\\ 0,\,\boldsymbol{b-1}\ ; & /\end{matrix}\right.\right)\#\,{}_pF_q(\boldsymbol{a};\,\boldsymbol{b};zA_+)\,{}_pF_q(\boldsymbol{a};\,\boldsymbol{b};z^*A_-)\#= \\ = \Gamma\left(\boldsymbol{b}/\boldsymbol{a}\right)\#\,{}_pF_q(\boldsymbol{a};\,\boldsymbol{b};A_+A_-)\# \tag{5. 12}$$

The angular integral is

$$\int_0^{2\pi}\frac{d\varphi}{2\pi}\#\,{}_pF_q(\boldsymbol{a};\,\boldsymbol{b};zA_+)\,{}_pF_q(\boldsymbol{a};\,\boldsymbol{b}\,;z^*A_-)\#=\#\,{}_{2p}F_{2q+1}(\boldsymbol{a}\,,\,\boldsymbol{a};1,\,\boldsymbol{b},\,\boldsymbol{b};A_+A_-|z|^2)\#= \\ = \left[\Gamma\left(\boldsymbol{b}/\boldsymbol{a}\right)\right]^2\#G_{2p,2q+2}^{1,2p}\left(-A_+A_-|z|^2\left|\begin{matrix}\boldsymbol{1-a},\ \boldsymbol{1-a}\,; & /\\ 0\,; & 0,\boldsymbol{1-b},\boldsymbol{1-b}\end{matrix}\right.\right)\# \tag{5.13}$$

so, particularly we obtain the following new integral in real space

$$\int_0^\infty d(|z|^2)G_{p,q+1}^{q+1,0}\left(|z|^2\left|\begin{matrix}/\,; & \boldsymbol{a-1}\\ 0,\,\boldsymbol{b-1}\ ; & /\end{matrix}\right.\right)\#G_{2p,2q+2}^{1,2p}\left(-A_+A_-|z|^2\left|\begin{matrix}\boldsymbol{1-a},\ \boldsymbol{1-a}\,; & /\\ 0\,; & 0,\boldsymbol{1-b},\boldsymbol{1-b}\end{matrix}\right.\right)\#= \\ = \#G_{p,\,q+1}^{1,\,p}\left(-A_+A_-\left|\begin{matrix}\boldsymbol{1-a}\,; & /\\ 0\,; & \boldsymbol{1-b}\end{matrix}\right.\right)\# \tag{5.14}$$

In the subsidiary, this can constitute another proof of the validity of the DOOT technique.

Since, according to the DOOT rules, the operators can be treated as simple c-numbers, the validity of the above integral can be checked using the properties of Meijer's G functions [Mathai, Saxena, 1973], [Popov, 2022].

The expectation value of an operator $\hat{O}$ in the BG-CSs representation is

$$<\hat{O}>_z\,\equiv\,<z|\hat{O}|z>=\frac{1}{{}_pF_q(\boldsymbol{a};\,\boldsymbol{b};|z|^2)}\sum_{n,n'}\frac{\left(z^*\right)^n(z')^{n'}}{\sqrt{\rho_{BG}(n)}\,\sqrt{\rho_{BG}(n')}}\ <n|\hat{O}|n'> \tag{5.15}$$

As an example, the expectation value of the product operator $A_+A_-$, is

$$< (A_+ A_-)^s >_z \equiv < z \,|\, (A_+ A_-)^s \,|\, z > = \frac{1}{{}_pF_q(\boldsymbol{a};\boldsymbol{b};z\,|^2)} \sum_{n=0}^{\infty} \frac{1}{\rho(n)} \left(|\,z\,|^2\right)^n \left[e(n)\right]^s =$$
$$= \frac{1}{{}_pF_q(\boldsymbol{a};\boldsymbol{b};z\,|^2)} \left[e(|\,z\,|^2\ \hat{D}_{|z|^2})\right]^s {}_pF_q(\boldsymbol{a};\boldsymbol{b};z\,|^2) \tag{5.16}$$

where $\hat{D}_{|z|^2} \equiv \frac{\partial}{\partial\,|\,z\,|^2}$ , $(|\,z\,|^2\ \hat{D}_{|z|^2})^s \equiv \left(|\,z\,|^2 \frac{\partial}{\partial\,|\,z\,|^2}\right)^s$ and we obtain that, under the sign of summation involving $n$, each $n$ from the expression of $e(n)$ must be replaced with $|\,z\,|^2\ \hat{D}_{|z|^2}$ .

Consequently, we can formulate the following *rule for BG-CSs of an arbitrary function*: if we calculate the expected values for a function depending on the ordered product $A_+ A_-$,, each product $A_+ A_-$ can be replaced with $|\,z\,|^2\ \hat{D}_{|z|^2}$ :

$$\# < f\left(A_+ A_-\right) >_z \# = f\left[e(|\,z\,|^2\ \hat{D}_{|z|^2})\right] \tag{5.17}$$

i.e. in the expectation value *in the BG-CSs representation the ordered product* $A_+ A_-$ *can be replaced by the variable* $|\,z\,|^2\ \hat{D}_{|z|^2}$ .

This rule may apply, e.g. to calculate *expected values for BG-CSs* of some oscillators with *linear energy spectra*, like HO-1D, or pseudoharmonic oscillator (PHO) [Popov, 2017 a], k-coherent states [Popov, 2017 b], or to some quantum oscillators with non-equidistant energy levels (Pöschl-Teller oscillators) [Popov, 2017 c].

For the integer powers of number operator $\hat{\boldsymbol{N}}$, which is diagonal in $\{|\,n>\}$- basis, we have

$$< \hat{\boldsymbol{N}}^s >_z^{(BG)} = \frac{1}{{}_pF_q(\boldsymbol{a};\boldsymbol{b};|\,z\,|^2)} \left(|\,z\,|^2\ \hat{D}_{|z|^2}\right)^s {}_pF_q(\boldsymbol{a};\boldsymbol{b};|\,z\,|^2) \tag{5.18}$$

Consequently, the Mandel parameter becomes

$$Q_{|z|}^{(BG)} = |\,z\,|^2 \left[ \frac{\left(\hat{D}_{|z|^2}\right)^2 {}_pF_q(\boldsymbol{a};\boldsymbol{b};|\,z\,|^2)}{\hat{D}_{|z|^2}\ {}_pF_q(\boldsymbol{a};\boldsymbol{b};|\,z\,|^2)} - \frac{\hat{D}_{|z|^2}\ {}_pF_q(\boldsymbol{a};\boldsymbol{b};|\,z\,|^2)}{{}_pF_q(\boldsymbol{a};\boldsymbol{b};|\,z\,|^2)} \right] \tag{5.19}$$

and its sign and value must be calculated for each individual case.

**Klauder-Perelomov nonlinear coherent states** (KP-NCSs) are defined by the action of a displacement operator on the vacuum Fock vector $|\,0>$, in the manner of Perelomov [Perelomov, 1972]. But, because the nonlinear operators $A_+$ and $A_-$ are not commutable, we have to use the pair of canonical operators $A_+$ and $\tilde{A}_-$ . So, the KP-NCSs are defined as:

*Definition 1*, with the help of the generalized displacement operator $\mathcal{D}(z)$:

$$|\,z>_{KP} = \mathcal{D}(z)\,|\,0> \tag{5.20}$$

$$| z >_{KP} = \frac{1}{\sqrt{\mathcal{N}_1(|z|^2)}} \exp(z A_+) \exp(-z^* \tilde{A}_-) | 0 > = \frac{1}{\sqrt{\mathcal{N}_1(|z|^2)}} \exp(z A_+) | 0 > =$$

$$= \frac{1}{\sqrt{\mathcal{N}_1(|z|^2)}} \sum_n \frac{z^n}{n!} (A_+)^n | 0 > = \frac{1}{\sqrt{\mathcal{N}_1(|z|^2)}} \sum_n \frac{z^n}{n!} \sqrt{\rho(n)} | n > = \tag{5.21}$$

$$= \frac{1}{\sqrt{{}_q F_p(\boldsymbol{b};\, \boldsymbol{a}; | z |^2)}} \sum_{n=0}^{\infty} \frac{z^n}{\sqrt{\rho_{KP}(n)}} | n >$$

and, after normalization,

$$\mathcal{N}_1(|z|^2) = \mathcal{N}_{KP}(| z |^2) = \sum_{n=0}^{\infty} \frac{(| z |^2)^n}{\rho_{KP}(n)} = \sum_{n=0}^{\infty} \frac{\prod_{j=1}^{q} (b_j)_n}{\prod_{i=1}^{p} (a_i)_n} \frac{(| z |^2)^n}{n!} = {}_q F_p(\boldsymbol{b};\, \boldsymbol{a}; | z |^2) . \tag{5.22}$$

With this result, the expansion of KP-NCSs becomes

$$| z >_{KP} = \frac{1}{\sqrt{{}_q F_p(\boldsymbol{b};\, \boldsymbol{a}; | z |^2)}} \sum_{n=0}^{\infty} \frac{z^n}{\sqrt{\rho_{KP}(n)}} | n > \tag{5.23}$$

The convergence radius $R_{KP}$ of the KP-NCSs is calculated similarly as for BG-CSs:

$$R_{KP} = \lim_{n\to\infty} \frac{\rho_{KP}(n)}{\rho_{KP}(n+1)} = \lim_{n\to\infty} \frac{(n!)^2}{\rho_{BG}(n)} \frac{\rho_{BG}(n+1)}{[(n+1)!]^2} = \lim_{n\to\infty} n^{q-p-1} = \begin{cases} \infty, & \text{if} \quad q-p-1>0 \\ 1\,, & \text{if} \quad q-p-1>0 \\ 0\,, & \text{if} \quad q-p-1>0 \end{cases} \tag{5.24}$$

The inversion of the *p* and *q* indices can be observed, compared to the case of BG-NCSs.

*Definition 2*, with the help of the annihilation operator $\tilde{A}_-$:

$$\tilde{A}_- | z >_{KP} = z | z >_{KP} \tag{5.25}$$

and their expansion according to the Fock vectors becomes identical to the one above, where $\rho_{KP}(n) = \tilde{\rho}(n) = \frac{(n!)^2}{\rho(n)}$. The proof is basically identical to that in *Definition 1* for BG-CSs, in addition taking into account the equality $\sqrt{\frac{\tilde{e}(n)}{\rho_{KP}(n)}} = \sqrt{\frac{1}{\rho_{KP}(n-1)}}$ and of course, after changing the summation index $n-1=m$, and eliminating the term with $m=-1$.

According to DOOT rules, the KP-NCSs projector is

$$| z >_{KP} \, {}_{KP}< z | = \frac{1}{{}_q F_p(\boldsymbol{b};\, \boldsymbol{a}; | z |^2)} \# \frac{{}_q F_p(\boldsymbol{b};\, \boldsymbol{a}; z\tilde{A}_+)\, {}_q F_p(\boldsymbol{b};\, \boldsymbol{a}; z^*\tilde{A}_-)}{{}_q F_p(\boldsymbol{b};\, \boldsymbol{a}; \tilde{A}_+\tilde{A}_-)} \# \tag{5.26}$$

where we used the fact that the vacuum projector can be expressed also through the product of pair operators $\tilde{A}_+\tilde{A}_-$:

$$\hat{I} = \sum_{n=0}^{\infty} |\, n >< n \,| = \sum_{n=0}^{\infty} \frac{1}{\rho_{KP}(n)} \left(\tilde{A}_+\right)^n |\, 0 >< 0 \,| \left(\tilde{A}_-\right)^n =$$
$$= |\, 0 >< 0 \,| \sum_{n=0}^{\infty} \frac{\#(\tilde{A}_+ \tilde{A}_-)^n \#}{\rho_{KP}(n)} = |\, 0 >< 0 \,| \#_q F_p(\boldsymbol{b};\, \boldsymbol{a};\, \tilde{A}_+ \tilde{A}_-)\# \qquad (5.27)$$

$$|\, 0 >< 0 \,| = \frac{1}{\#_q F_p(\boldsymbol{b};\, \boldsymbol{a};\, \tilde{A}_+ \tilde{A}_-)\#} \qquad (5.28)$$

Consequently, the resolution of unity becomes

$$\hat{I} = \int d_{KP}(z)\, |\, z >_{KP\ \ KP}< z \,| =$$
$$= \sum_{n,n'=0}^{\infty} \frac{|\, n >< n' \,|}{\sqrt{\rho_{KP}(n)}\, \sqrt{\rho_{KP}(n')}} \int_0^{R_{KP}} d(|\, z \,|^2) \frac{h_{KP}(|\, z \,|^2)}{{}_q F_p(\boldsymbol{b};\, \boldsymbol{a};|\, z \,|^2)} \int_0^{2\pi} \frac{d\varphi}{2\pi} (z^*)^n\, (z')^{n'} \qquad (5.29)$$

The angular integral above is $(|\, z \,|^2)\, \delta_{nn'}$ and with $\tilde{h}_{KP}(|\, z \,|^2) \equiv h_{KP}(|\, z \,|^2)\ /\ {}_q F_p(\boldsymbol{b};\, \boldsymbol{a};|\, z \,|^2)$ and $n = s - 1$, the corresponding moment problem is

$$\int_0^{R_{KP}} d(|\, z \,|^2)\, \tilde{h}_{KP}(|\, z \,|^2)(|\, z \,|^2)^{s-1} = \rho_{KP}(s-1) = \Gamma\left(\boldsymbol{b}\ /\, \boldsymbol{a}\right) \frac{\Gamma(s) \prod_{i=1}^{p} \Gamma(a_i - 1 + s)}{\prod_{j=1}^{q} \Gamma(b_j - 1 + s)} \qquad (5.30)$$

whose solution is also a G-Meijer function, so that, finally, the integration measure becomes

$$d_{KP}(z) = \Gamma\left(\boldsymbol{b}\ /\, \boldsymbol{a}\right) \frac{d\varphi}{2\pi} d(|\, z \,|^2)\, G_{q,p+1}^{p+1,0}\left(|\, z \,|^2 \left|\, \begin{matrix} / \,; & \boldsymbol{b-1} \\ 0, \boldsymbol{a}-1 \,; & / \end{matrix} \right.\right) {}_q F_p(\boldsymbol{b};\, \boldsymbol{a};|\, z \,|^2) \qquad (5.31)$$

Substituting this expression in the unity decomposition relation, and performing the angular integral

$$\int_0^{2\pi} \frac{d\varphi}{2\pi} \#_q F_p(\boldsymbol{b};\, \boldsymbol{a};\, z\tilde{A}_+)\, {}_q F_p(\boldsymbol{b};\, \boldsymbol{a}\,;\, z^* \tilde{A}_-)\# = \#_{2p} F_{2q+1}(\boldsymbol{b}\,,\, \boldsymbol{b};\, 1,\, \boldsymbol{a},\, \boldsymbol{a};\, \tilde{A}_+ \tilde{A}_- \,|\, z \,|^2)\# =$$
$$= \left[\Gamma\left(\boldsymbol{a}/\boldsymbol{b}\right)\right]^2 \# G_{2q,2p+2}^{1,2q}\left(-\tilde{A}_+ \tilde{A}_- \,|\, z \,|^2 \left|\, \begin{matrix} \boldsymbol{1-b},\, \boldsymbol{1-b}\,; & / \\ 0\,; & 0,\ \boldsymbol{1-a}, \boldsymbol{1-a} \end{matrix} \right.\right)\# \qquad (5.32)$$

after a few simple operations we arrive at a new integral, in the real space, the validity of which can be verified by using the properties of the G-Meijer functions:

$$\int_0^{R_{KP}} d(|\, z \,|^2)\, G_{q,p+1}^{p+1,0}\left(|\, z \,|^2 \left|\, \begin{matrix} / \,; & \boldsymbol{b-1} \\ 0, \boldsymbol{a}-1 \,; & / \end{matrix} \right.\right) \# G_{2q,2p+2}^{1,2q}\left(-\tilde{A}_+ \tilde{A}_- \,|\, z \,|^2 \left|\, \begin{matrix} \boldsymbol{1-b},\, \boldsymbol{1-b}\,; & / \\ 0\,; & 0,\ \boldsymbol{1-a}, \boldsymbol{1-a} \end{matrix} \right.\right)\# =$$
$$= G_{q,p+1}^{1,q}\left(-\tilde{A}_+ \tilde{A} \left|\, \begin{matrix} \boldsymbol{1-b}\,; & / \\ 0\,; & \boldsymbol{1-a} \end{matrix} \right.\right) = \Gamma\left(\boldsymbol{a}/\boldsymbol{b}\right)\, {}_q F_p(\boldsymbol{b};\, \boldsymbol{a};\, \tilde{A}_+ \tilde{A})$$

(5.33)

We remind that, according to the DOOT rules, the operators can be treated as simple c-numbers, so they can be replaced with some arbitrary constants.

### 6. Similarities between BG- and KP-CSs

As a partial conclusion, we can say that the pairs of dual operators $A_-$ and $\tilde{A}_+$ (through $\tilde{\mathcal{D}}(z)$) generates the NCSs in the Barut-Girardello sense, $|z>_{BG}$, while the pair of operators $A_+$ (through $\mathcal{D}(z)$) and $\tilde{A}_-$ acts in such a way that it generates a NCSs of the Klauder-Perelomov type, $|z>_{KP}$.

Therefore, the following operator links - NCSs result:

| | ***Definition 1*** | ***Definition 2*** |
|---|---|---|
| BG-CSs | $A_- \quad \Leftrightarrow \quad \|z>_{BG}$ | $\tilde{A}_+ \to \tilde{\mathcal{D}}(z) \quad \Leftrightarrow \quad \|z>_{BG}$ |
| KP-CSs | $A_+ \to \mathcal{D}(z) \quad \Leftrightarrow \quad \|z>_{KP}$ | $\tilde{A}_- \quad \Leftrightarrow \quad \|z>_{KP}$ |

From the properties of the pairs of dual operators $\left(A_-\ ,\ A_+\right)$ and $\left(\tilde{A}_-\ ,\ \tilde{A}_+\right)$, i.e.

$$< n|\begin{Bmatrix} A_-\tilde{A}_+ \\ \tilde{A}_+A_- \end{Bmatrix}|n>=n+1 \qquad , \qquad < n|\begin{Bmatrix} \tilde{A}_+A_- \\ A_+\tilde{A}_- \end{Bmatrix}|n>=n \tag{6.1}$$

the following average values in the BG- , respectively KP-CSs representations ${}_{\begin{Bmatrix} BG \\ KP \end{Bmatrix}}< z|...|z>_{\begin{Bmatrix} BG \\ KP \end{Bmatrix}} \equiv <...>_{\begin{Bmatrix} BG \\ KP \end{Bmatrix}}$ can be deduced:

$$\begin{aligned} <\begin{Bmatrix} \tilde{A}_+A_- \\ A_+\tilde{A}_- \end{Bmatrix}>_{\begin{Bmatrix} BG \\ KP \end{Bmatrix}} &= \frac{1}{\mathcal{N}_{\begin{Bmatrix} BG \\ KP \end{Bmatrix}}\left(|z|^2\right)}\sum_{n=0}^{\infty}\frac{\left(|z|^2\right)^n}{\rho_{\begin{Bmatrix} BG \\ KP \end{Bmatrix}}(n)} < n|\begin{Bmatrix} \tilde{A}_+A_- \\ A_+\tilde{A}_- \end{Bmatrix}|n>= \\ &= \frac{1}{\mathcal{N}_{\begin{Bmatrix} BG \\ KP \end{Bmatrix}}\left(|z|^2\right)}\sum_{n=0}^{\infty}\frac{\left(|z|^2\right)^n}{\rho_{\begin{Bmatrix} BG \\ KP \end{Bmatrix}}(n)}n = \frac{1}{\mathcal{N}_{\begin{Bmatrix} BG \\ KP \end{Bmatrix}}\left(|z|^2\right)}\left(|z|^2\frac{d}{d|z|^2}\right)\mathcal{N}_{\begin{Bmatrix} BG \\ KP \end{Bmatrix}}\left(|z|^2\right) \end{aligned} \tag{6.2}$$

i.e., the average values of the pairs of dual operators in the two representations are dual.

By generalization, the average values of any function that depends on these products ordered by operators are equal:

$$< F\left(\begin{Bmatrix} \tilde{A}_+A_- \\ A_+\tilde{A}_- \end{Bmatrix}\right)>_{\begin{Bmatrix} BG \\ KP \end{Bmatrix}} = \frac{1}{\mathcal{N}_{\begin{Bmatrix} BG \\ KP \end{Bmatrix}}\left(|z|^2\right)}F\left(|z|^2\frac{d}{d|z|^2}\right)\mathcal{N}_{\begin{Bmatrix} BG \\ KP \end{Bmatrix}}\left(|z|^2\right) \tag{6.3}$$

which means that, in the calculation of the average value in the NCSs representation, each ordered product $\#\left\{\begin{matrix}\tilde{A}_+ A_- \\ A_+ \tilde{A}_-\end{matrix}\right\}\#$ will be replaced by the derivative $|z|^2 \dfrac{d}{d|z|^2}$ of the normalization function $\mathcal{N}_{\left\{\begin{matrix}BG\\KP\end{matrix}\right\}}\left(|z|^2\right)$.

For example, the average value of the Hamiltonian $\hat{\mathcal{H}}\,|n>=e(n)\,|n>$, in the NCSs representation will be

$$<\#\hat{\mathcal{H}}\#>_{\left\{\begin{matrix}BG\\KP\end{matrix}\right\}} = \frac{1}{\mathcal{N}_{\left\{\begin{matrix}BG\\KP\end{matrix}\right\}}\left(|z|^2\right)}\left\{\begin{matrix} e\left(|z|^2 \dfrac{d}{d|z|^2}\right) \\ \tilde{e}\left(|z|^2 \dfrac{d}{d|z|^2}\right)\end{matrix}\right\}\mathcal{N}_{\left\{\begin{matrix}BG\\KP\end{matrix}\right\}}\left(|z|^2\right) \tag{6.4}$$

and similarly, of an integer power of the operator number of particles $\hat{\mathcal{N}}^s$ :

$$<\hat{\mathcal{N}}^s>_{\left\{\begin{matrix}BG\\KP\end{matrix}\right\}} = \frac{1}{\mathcal{N}_{\left\{\begin{matrix}BG\\KP\end{matrix}\right\}}\left(|z|^2\right)}\left(|z|^2 \frac{d}{d|z|^2}\right)^s \mathcal{N}_{\left\{\begin{matrix}BG\\KP\end{matrix}\right\}}\left(|z|^2\right) \tag{6.5}$$

Consequently, the Mandel parameter in the NCSs representation reads

$$Q_{|z|}^{\left\{\begin{matrix}BG\\KP\end{matrix}\right\}} = |z|^2\left[\frac{\left(D_{|z|^2}\right)^2 \mathcal{N}_{\left\{\begin{matrix}BG\\KP\end{matrix}\right\}}\left(|z|^2\right)}{D_{|z|^2}\mathcal{N}_{\left\{\begin{matrix}BG\\KP\end{matrix}\right\}}\left(|z|^2\right)} - \frac{D_{|z|^2}\mathcal{N}_{\left\{\begin{matrix}BG\\KP\end{matrix}\right\}}\left(|z|^2\right)}{\mathcal{N}_{\left\{\begin{matrix}BG\\KP\end{matrix}\right\}}\left(|z|^2\right)}\right] \tag{6.6}$$

Starting from the general expression of the density operator, let's see what are the concrete expressions for the distribution functions *Q*- and *P*-, generated by the mixed operators $A_\pm$ and $\tilde{A}_\pm$. Because the expression of the density operator is essentially determined by the expression of the energy eigenvalues, we will specifically refer to the systems with linear energy spectra, $E_n = \hbar\omega\left(n + e_0\right)$.

*For the pair* $\left(A_+\ ,\ A_-\right)$:

$$\begin{aligned} \hat{\rho}_{BG} &= \frac{1}{\bar{n}+1}\#\frac{1}{\mathcal{N}_{BG}\left(A_+ A_-\right)}\mathcal{N}_{BG}\left(\frac{\bar{n}}{\bar{n}+1}A_+ A_-\right)\#= \\ &=\#\frac{1}{\mathcal{N}_{BG}\left(A_+ A_-\right)}\int d\mu_{BG}(z)\,P_{BG}(|z|^2)\frac{1}{\mathcal{N}_{BG}\left(|z|^2\right)}\sum_{n,n'=0}^{\infty}\frac{\left(zA_+\right)^n}{\rho_{BG}(n)}\frac{\left(z^* A_-\right)^{n'}}{\rho_{BG}(n')}\#= \end{aligned} \tag{6.7}$$

After angular integration we must have

$$\frac{1}{\bar{n}+1}\#\mathcal{N}_{BG}\left(\frac{\bar{n}}{\bar{n}+1}A_+A_-\right)\#=$$

$$=\Gamma(\boldsymbol{a}/\boldsymbol{b})\sum_{n=0}^{\infty}\frac{(A_+A_-)^n}{[\rho_{BG}(n)]^2}\#\int_0^{R_{BG}}d(|z|^2)\,G_{p,\,q+1}^{q+1,0}\left(|z|^2\left|\begin{matrix}/\,; & \boldsymbol{a-1}\\ 0\,,\boldsymbol{b-1}\,; & /\end{matrix}\right.\right)P_{BG}(|z|^2)(|z|^2)^n \tag{6.8}$$

Equality is valid if the *P*-quasi distribution function has the expression:

$$P_{BG}(|z|^2)=\frac{1}{\bar{n}}\frac{G_{p,\,q+1}^{q+1,0}\left(\frac{\bar{n}+1}{\bar{n}}|z|^2\left|\begin{matrix}/\,; & \boldsymbol{a-1}\\ 0\,,\boldsymbol{b-1}\,; & /\end{matrix}\right.\right)}{G_{p,\,q+1}^{q+1,0}\left(|z|^2\left|\begin{matrix}/\,; & \boldsymbol{a-1}\\ 0\,,\boldsymbol{b-1}\,; & /\end{matrix}\right.\right)} \tag{6.9}$$

*For the pair* $(\tilde{A}_+\,,\ \tilde{A}_-)$:

$$\hat{\rho}_{KP}=\frac{1}{\bar{n}+1}\#\frac{1}{\mathcal{N}_{KP}(\tilde{A}_+\tilde{A}_-)}\mathcal{N}_{KP}\left(\frac{\bar{n}}{\bar{n}+1}\tilde{A}_+\tilde{A}_-\right)\#=$$

$$=\#\frac{1}{\mathcal{N}_{KP}(\tilde{A}_+\tilde{A}_-)}\int d\mu_{KP}(z)\,P_{KP}(|z|^2)\frac{1}{\mathcal{N}_{KP}(|z|^2)}\sum_{n,n'=0}^{\infty}\frac{(z\tilde{A}_+)^n}{\rho_{KP}(n)}\frac{(z^*\tilde{A}_-)^{n'}}{\rho_{KP}(n')}\# \tag{6.10}$$

After similar calculations, we obtain that the *P*-quasi distribution function is

$$P_{KP}(|z|^2)=\frac{1}{\bar{n}}\frac{G_{q,\,p+1}^{p+1,0}\left(\frac{\bar{n}+1}{\bar{n}}|z|^2\left|\begin{matrix}/\,; & \boldsymbol{b-1}\\ 0\,,\boldsymbol{a-1}\,; & /\end{matrix}\right.\right)}{G_{q,\,p+1}^{p+1,0}\left(|z|^2\left|\begin{matrix}/\,; & \boldsymbol{b-1}\\ 0\,,\boldsymbol{a-1}\,; & /\end{matrix}\right.\right)} \tag{6.11}$$

The $Q$ – distribution function (also called Husimi’s distribution function), defined as the expectation value of the density operator in the CSs representation, can be written as

$$<Q>_z^{\left\{\begin{smallmatrix}BG\\KP\end{smallmatrix}\right\}}\equiv{}_{\left\{\begin{smallmatrix}BG\\KP\end{smallmatrix}\right\}}<z\,|\,\hat{\rho}\,|\,z>_{\left\{\begin{smallmatrix}BG\\KP\end{smallmatrix}\right\}}=\frac{1}{\bar{n}+1}\frac{1}{\mathcal{N}_{\left\{\begin{smallmatrix}BG\\KP\end{smallmatrix}\right\}}(|z|^2)}\mathcal{N}_{\left\{\begin{smallmatrix}BG\\KP\end{smallmatrix}\right\}}\left(\frac{\bar{n}}{\bar{n}+1}|z|^2\right) \tag{6.12}$$

The thermal counterpart of the Mandel parameter being the same for both type of NCS, BG- and KP-, we have examined in the previous section so we will not repeat the calculations again here.

Let's see now what is the connection between the dual BG- and KP-NCSs. This problem was examined for the first time by Ali et al. [Ali, 2004], [Roknizadeh, Tavassolly, 2004]. Similar to the idea from the paper of Ali et al. [Ali, 2004], but with some small differences, let's introduce a transformation operator and their inverse which makes the jump from one type of CSs to another

$$\mathcal{T}_{p,q}(|z|^2)\equiv\sqrt{\frac{{}_pF_q(\boldsymbol{a};\,\boldsymbol{b};|z|^2)}{{}_qF_p(\boldsymbol{b};\,\boldsymbol{a};|z|^2)}}\sum_n\sqrt{\frac{\rho_{BG}(n)}{\rho_{KP}(n)}}\,|\,n><n\,| \tag{6.13}$$

$$\mathcal{T}^{-1}_{p,q}(|z|^2) \equiv \sqrt{\frac{{}_qF_p(\boldsymbol{b};\boldsymbol{a};|z|^2)}{{}_pF_q(\boldsymbol{a};\boldsymbol{b};|z|^2)}} \sum_n \sqrt{\frac{\rho_{KP}(n)}{\rho_{BG}(n)}} |n><n| \tag{6.14}$$

which follows due to the completion relation $\sum_n |n><n| = 1$ .

These operators transform Barut-Girardello type NCSs into Klauder-Perelomov type NCSs and vice versa, connecting the two dual types of CSs:

$$\mathcal{T}_{p,q}(|z|^2)|z>_{BG} = |z>_{KP} \qquad , \qquad \mathcal{T}^{-1}_{p,q}(|z|^2)|z>_{KP} = |z>_{BG} \tag{6.15}$$

The action of these operators on the pair of NCSs BG- and KP- can be written in matrix form as follows:

$$\begin{pmatrix} \mathcal{T}_{p,q}(|z|^2) & 0 \\ 0 & \mathcal{T}^{-1}_{p,q}(|z|^2) \end{pmatrix} \begin{pmatrix} |z>_{BG} \\ |z>_{KP} \end{pmatrix} = \begin{pmatrix} |z>_{KP} \\ |z>_{BG} \end{pmatrix} \tag{6.16}$$

So, the operator matrix has the same action as the Pauli $\hat{X}$ gate operator (NOT gate) $\hat{X} = \begin{pmatrix} 0 & 1 \\ 1 & 0 \end{pmatrix}$:

$$\begin{pmatrix} 0 & 1 \\ 1 & 0 \end{pmatrix} \begin{pmatrix} |z>_{BG} \\ |z>_{KP} \end{pmatrix} = \begin{pmatrix} |z>_{KP} \\ |z>_{BG} \end{pmatrix} \tag{6.17}$$

Roknizadeh and Tavassoly [Roknizadeh, 2004] used similar operators, which they applied to canonical coherent states (for HO-1D), and obtain (up to the normalization function) the dual BG- and KP-NCSs.

Finally, let's point out another result of the formalism introduced by Roy and Roy [Roy and Roy, 2000], [Roknizadeh, 2004]: the two types of dual NCSs, BG- and KP-, can be built with the help of a pair of displacement operators $\tilde{\mathcal{D}}(z)$ and $\mathcal{D}(z)$ whose component includes the mixed generators $A_-$ and $\tilde{A}_+$, respectively $\tilde{A}_-$ and $A_+$ .

In a matrix form, these actions are

$$\begin{pmatrix} A_- \\ \tilde{A}_- \end{pmatrix} \begin{pmatrix} |z>_{BG} & 0 \\ 0 & |z>_{KP} \end{pmatrix} = \begin{pmatrix} |z>_{BG} \\ |z>_{KP} \end{pmatrix} \tag{6.18}$$

$$\begin{pmatrix} |z>_{BG} \\ |z>_{KP} \end{pmatrix} = \begin{pmatrix} \mathcal{D}_2(z) \\ \mathcal{D}_1(z) \end{pmatrix} \begin{pmatrix} |0> & 0 \\ 0 & |0> \end{pmatrix} \tag{6.19}$$

In this manner we highlighted different ways of building the two dual types of NCSs, BG- and KP-CSs.

## 7. Some concluding remarks

In the paper we have pointed out some new manners to build two kinds of generalized or nonlinear coherent states (Barut-Girardello and Klauder-Perelomov), and examined their statistical properties for pure and mixed (thermal) states.

In conclusion, the following remarks can be made:

- The "tilde" operation has the same effect as simple complex conjugation. It does not change the order of operators in a product.
- The pairs of operators $\left(\tilde{A}_+ ,\ A_-\right)$, respectively $\left(A_+ ,\ \tilde{A}_-\right)$ are dual, in the sense that they are responsible for the construction of BG-CSs, respectively KP-CSs.
- The duality between BG-CSs and respectively KP-CSs is manifested through the reciprocal interchanges of entities contained in the following table::

| ***Entity*** | ***BG-CSs*** | ***KP-CSs*** |
|---|---|---|
| Dual generating operators | $\left(\tilde{A}_+ ,\ A_-\right)$ | $\left(A_+ ,\ \tilde{A}_-\right)$ |
| Displacement operators | $\tilde{\mathcal{D}}(z)$ | $\mathcal{D}(z)$ |
| Interchanges of indices | $p \quad ; \quad q$ | $q \quad ; \quad p$ |
| Interchanges of real numbers | $\boldsymbol{a}\ ;\ \boldsymbol{b}$ | $\boldsymbol{b}\ ;\ \boldsymbol{a}$ |
| Structure functions | $\rho(n) \equiv \rho_{BG}(n)$ | $\tilde{\rho}(n) \equiv \rho_{KP}(n)$ |
| Normalization function | $\mathcal{N}_{BG}\left(A_+ A_-\right) = {}_pF_q(\boldsymbol{a};\boldsymbol{b};z\,\vert^2)$ | $\mathcal{N}_{KP}\left(\vert z\vert^2\right) = {}_qF_p(\boldsymbol{b};\boldsymbol{a};\vert z\vert^2)$ |
| Vacuum projector $\vert 0><0\vert$ | $\left[\#_pF_q(\boldsymbol{a};\boldsymbol{b};A_+A_-)\#\right]^{-1}$ | $\left[\#_qF_p(\boldsymbol{b};\boldsymbol{a};\tilde{A}_+\tilde{A})\#\right]^{-1}$ |
| Density operator | $\hat{\rho} = \hat{\rho}_{BG} = f\left(A_+ A_-\right)$ | $\hat{\rho} = \hat{\rho}_{KP} = f\left(\tilde{A}_+ \tilde{A}_-\right)$ |
| Husimi's function | $<Q>_z^{BG} = f\left(A_+ A_-\right)$ | $<Q>_z^{KP} = f\left(\tilde{A}_+ \tilde{A}_-\right)$ |
| *P*-distribution function | $P_{BG}(\vert z\vert^2) = f\left[G_{p,\ q+1}^{q+1,0}\left(...\vert\ ...\right)\right]$ | $P_{KP}(\vert z\vert^2) = f\left[G_{q,\ p+1}^{p+1,0}\left(...\vert\ ...\right)\right]$ |
| Mandel parameter | $Q_{\vert z\vert}^{BG} = f\left[\mathcal{N}_{BG}\left(\vert z\vert^2\right)\right]$ | $Q_{\vert z\vert}^{KP} = f\left[\mathcal{N}_{KP}\left(\vert z\vert^2\right)\right]$ |
| Jump operators | $\mathcal{T}_{p,q}(\vert z\vert^2)$ | $\mathcal{T}_{p,q}^{-1}(\vert z\vert^2)$ |

In conclusion, the following points can be made: In general, for the generalized or nonlinear coherent states (NCSs), the BG- and, respectively, KP- type coherent states are not identical, as in the case of the one-dimensional harmonic oscillator (HO-1D). However, analogies or dualities can be found between the two types. These dualities can be revealed by introducing some dual operators that can be grouped into pairs that form canonical operators (their switch is equal to the unity operator). These pairs of dual operators allow the construction of generalized displacement operators. These pairs of dual operators are responsible for building and revealing the properties of the two types of NCSs, so that each type of NCSs can be constructed through two different definitions, involving dual operators.